\documentclass[a4paper,12pt]{article}
\usepackage[T1]{fontenc}
\usepackage[latin5]{inputenc}
\begin{document}
\begin{center}
\huge
Superposition Principle and Young Type Double-Slit Experiment in Vacuum
\huge
\end{center}
\noindent
Ali Savaş \\
Sarmaşık Mah., Ilıca, Fatsa, Ordu, Turkey \\
\noindent
\newline
In this study, it is shown with reasons that superposition principle does not work in vacuum. This case can be observed by Young type double slit experiment to be carried out.  Since field-field interaction is carried through charged particles, in the absence of charged particles linear superposition of two fields is not possible and interference will not be observed.  

\section{Introduction}
\noindent
As it is known, the electromagnetic field is expressed by linear equations both in classical and quantum mechanics. Nevertheless, the idea of necessity 
of expressing the electromagnetic field by a nonlinear equation has always kept the physicists mind busy.  Classically, the nonlinear theory of Born and 
Infeld$^{1}$ is the most known among these. Although no evidence of necessity of the non-linearity of the electromagnetic field was obtained classically, 
in quantum mechanics this situation is a little different. Due to uncertainty principle, photon-photon interaction$^{ 2}$(which is known as Delbrück scattering) is possible by the way that the two photons vanish by creating an electron-positron pair and afterwards annihilation of the electron-positron 
pair to create a different photon pair. In this case, the electric and magnetic permeability tensor of vacuum$^{2}$ -carrying information about the electron-positron pair-   will not be equal to 1, and since the electric and magnetic field will slowly change.

\begin{equation}
 \bf{D}_{i} = \sum_k \epsilon_{ik}\bf{E}_{k},      \bf{B}_{i} =\sum_k\mu_{ik}\bf{H}_{k}
\end{equation}

where
\begin{equation}
 \epsilon_{ik} = \delta_{ik}+(e\hbar /45\pi m^{4}c^{7})[2(E^{2}-B^{2})\delta_{ik}+7B_{i}B_{k}]+\cdots
\end{equation}

\begin{equation}
 \mu_{ik} = \delta_{ik}+(e\hbar /45\pi m^{4}c^{7})[2(B^{2}-E^{2})\delta_{ik}+7E_{i}E_{k}]+\cdots
\end{equation}
\noindent
Here e and m are the charge and mass of electron. This result was obtained first by Euler and Kockel$^{3}$.\newline
As two things of same character would not interact directly, information only could be obtained indirectly in nature.  That is, we can obtain information about charged particles through field and about field through charged particles. Since field-field interaction is through charged particles, in the absence of charged particles we cannot obtain information about a field through the other field. On the account of uncertainty principle, since simultaneous measurements on both field and charged particles is not possible, definite information about one would destroy information about the other. Hence, while using information about field information about charged particle will not appear in this expression. However, while information loss due to uncertainty principle is taken into account, linear superposition of two fields in vacuum would not be possible for the extra terms appearing. 
The objective of this study is to show that linear superposition principle does not work in vacuum. For this purpose, it will be shown both in classical and quantum electrodynamics (QED) aspects that linear superposition principle will not work for the Young type double-slit experiment to be performed in vacuum and hence interference pattern will not be observed.

\section{Classical Description of Interference}

According to the classical theory, interference occurs from two electromagnetic fields passing through double-slit.\\

\noindent
Linearly polarized monochromatic plane waves passing through each slit can be expressed as

\begin{equation}
\bf{E}_{i}(\bf{r},t) = E_{i}e^{i(\bf{k}_{i}\bf{r}-\omega_{i}t-\Phi_{i})},     (i =1,2)
\end{equation}
\noindent
Here, E$_{i}$ is the real amplitude, $\bf{k_{i}}$ the wave vector, $\omega_{i}$ the cirrular frequency and $\Phi_{i}$ the phase. 
In this case, the electric field at point $(\bf{r},t)$ is expressed as the linear superposition of two fields,  

\begin{equation}
\bf{E}(\bf{r},t) = \bf{E}_{1}(\bf{r_{1}},t)+\bf{E}_{2}(\bf{r_{2}},t)
\end{equation}
\noindent
and intensity is the square of $\bf{E}(\bf{r},t)$

\begin{equation}
 I (\bf{r},t) = E_{1}^{2} + E_{2}^{2} + 2 E_{1} E_{2} cos [( \bf{k}_{2}-\bf{k}_{1}) \bf{r} - (\omega_{2}-\omega_{2}) t 
- (\Phi_{2}-\Phi_{1})]  
\end{equation}
\noindent
Classically, the book of Born and Wolf$^{4}$ may be looked at for information on theory and experiment.    

\section{Quantum Mechanical Description}
\noindent
In QED, the electromagnetic field is obtained by the quantization of classical electromagnetic field$^{5}$. In that case, the electromagnetic
field may be written as a sum of two parts containing positive and negative frequencies.

\begin{equation}
\bf{E}(\bf{r},t) = \bf{E}^{(+)}(\bf{r},t)+\bf{E}^{(-)}(\bf{r},t)
\end{equation}

\begin{equation}
\bf{E}^{(+)}(\bf{r},t) = [\bf{E}^{(+)}(\bf{r},t)]^{\dag}
\end{equation}
\noindent                                                                  
Field passing through two slits may be written as 

\begin{equation}
\bf{E}^{(+)}(\bf{r},t) = f(\bf{r},t)(a_{1}e^{i\bf{k}_{1}\bf{r}_{1}}+a_{2}e^{i\bf{k}_{2}\bf{r}_{2}})
\end{equation}

\noindent 
Here, $a_{1}$ and $a_{2}$ are creation operator and 
                                            
\begin{equation}
 f(\bf{r},t) = i(\hbar \omega /2)^{1/2} (\hat{e}_{k}/4\pi R)^{1/2}e^{-i\omega t}/r
\end{equation}
\noindent
R is the normalized volume radius and $\hat{e}_{k}$ is the unit polarization vector. In that case, intensity is in 
the following form.

\begin{equation}
 I(\bf{r},t) = Tr (\rho E^{(-)}(\bf{r},t) E^{(+)}(\bf{r},t)
\end{equation}
\noindent
Here, $\rho$ is density operator expressing field state. Inserting (2.5) (2.6) and (2.7) into (2.8), 

\begin{equation}
 I(\bf{r},t) = N [ Tr (\rho a_{1}^{\dag}a_{1}) + Tr (\rho a_{2}^{\dag}a_{2}) + 2 Tr ( a_{1}^{\dag}a_{2}) cos \Phi ]
\end{equation}
\noindent
will be obtained. Here, 

\begin{equation}
Tr (\rho a_{1}^{\dag}a_{2}) = Tr (\rho a_{1}^{\dag}a_{2}) e^{i\phi}
\end{equation}

\begin{equation}
N= \arrowvert f(\bf{r},t)\arrowvert^{2}
\end{equation}

\begin{equation}
\Phi= \bf{k}(\bf{r}_{1}-\bf{r}_{2}) +\phi
\end{equation}
\noindent
a$_{1}^{\dag}$  and  a$_{2}^{\dag}$ is the annihilation operator and $\phi$ is the phase.

\section{IV. Young Type Double-Slit Experiment in Vacuum}
\noindent
Classically, in the Young type double-slit experiment, for the two fields to be able make interference necessitates that the field passing through 
one slit must know the geometry of the field passing through another slit, and vice versa.  In that case, it is necessary that the fields passing 
through two slit to interact. Since in vacuum field-field interaction is not possible, the linear superposition principle will not work 
and hence interference will not be observed. 
\noindent
In QED, even though not present in literature yet, it is being dwelled upon the idea of the Dirac's$^{6}$ statement: ``Each photon interferes only with itself. Interference between different photons never occurs''  The second part of this statement has been proven to be wrong by 
Forrester, Godmundson and Johnson$^{7}$ and as well as by Brown and Twiss$^{8}$ by showing that photons emitted by different atoms 
exhibit interference. Taylor$^{9}$, Janossy and Naray$^{10}$ showed that interference was obtained by sending single photons. Lets consider 
an electron isolated from its surroundings. As the electron interacts with a photon, it will carry the information belonging to the photon any more.
And as the electron interacts with a second photon, it will transfer the information belonging the first photon to the second one.  From the physical 
point of view, there is no any difference whether the interaction takes place simultaneously or at different times except time. For that reason, in the double-slit experiment to be performed by sending single photons in the presence of charged particles, the photon passing through a slit will identify 
the geometry of the other slit through charged particles and interference will be possible in that way. In the case of performing the Young type 
double-slit experiment in vacuum by sending single photons, since there will be no interaction between the photons, interference will not occur. 
As a consequence, interference between different photons is possible in the presence of charged particles. Since the superposition principle will 
not work in vacuum, if the contribution resulting from uncertainty principle is ignored, the intensity will be equal to the sum of intensities of the fields passing through two slit. Hence, the intensity expressions in (2.3) and (2.9) will be of the form:

\begin{equation}
 I(\bf{r},t) =  E_{1}^{2}+ E_{2}^{2}
\end{equation}

\begin{equation}
 I(\bf{r},t) = N [ Tr (\rho a_{1}^{\dag}a_{1}) + Tr (\rho a_{2}^{\dag}a_{2}) ]
\end{equation}

\noindent
Since wave will not pass through vacuum, particles will pass. In that case, what to be observed in the Young type double-slit experiment in vacuum 
will be the particle characteristics of light.  
Furthermore, if the information of the firstly passing photons is lost, in the double-slit experiment carried out by sending single photons, due to external effects or due to interactions between the charged particles present in the environment still interference will not occur. This case was demonstrated through an experiment carried out by Pfleegor and Mandel$^{11}$ by the use of two independent same type laser sources. They stated that when the time interval between the sent times of photons is increased it was not possible to observe interference. 

\section{Discussion}
\noindent
Since in nature, information could be obtained only by indirect means it will not be possible for a photon to make interference with itself. Interference between different photons is possible only in the presence of charged particles. In that case, it is necessary that both statements of Dirac's to be false. This case can be observed by a Young type double-slit experiment in vacuum. For the reasons explained above, linear superposition of two fields in vacuum will not be possible and hence interference will not be observed. 
\section{Referances}
1. M. Born and L. Infeld, Proc. Roy. Soc. A144, 425(1934) \newline
2. J. M. Jauch and F Rohrlich, The Theory of Photons and Electrons, Addison-Wesley (1955) \newline
3. H. Euler and B. Kockel, Naturwiss. 23, 246 (1935)   \newline 
4. M. Born ve E. Wolt , Principles of Optics , $3^{rd}$  Edition, Pergamon , New York , (1965) \newline
5. D.F. Walls , American Journal of Physics 45 , 10 , (1977) \newline
6. P.A.M. Diraç, The Principles of Quantum Mechanics, $3^{rd}$  Edition, Clarendon Press,  Oxford, (1947) \newline
7. A.T. Forrester , R.A. Gudmundson and P.O. Johnson , Phys. Rev. 99 , 1691 , (1955) \newline
8. R.H. Brown , R.Q. Twiss , Nature 177 , 27 , (1956) \newline
9. I. Taylor , Proc. Cambridge Philos. 15 , 144 , (1909) \newline
10. L .Janossy , Z. Naroy , Nuovo Cimento 9 , 588 , (1958) \newline
11. L. Pfleegor , L. Mandel , Phys. Rew. 159 , 1084 , (1967) 

\end{document}